\documentclass[aps,prl,twocolumn,superscriptaddress]{revtex4}
\usepackage{graphicx}
\usepackage{epsfig}
\usepackage{color}

\def\vec#1{{\mathbf{#1}}}

\definecolor{red}{rgb}{1.000,0.000,0.000}

\begin{document}

\title{A model free energy for glasses}
\author{H. Frielinghaus}
\affiliation{Forschungszentrum J\"ulich GmbH, J\"ulich Centre of Neutron Science at Heinz Maier-Leibnitz Zentrum,  Lichtenbergstr.\ 1, D-85747 Garching, Germany}

\date{\today}

\begin{abstract}
\noindent
We develop a model free energy from an expansion that basically includes graphs without loops.
From this calculation, we derive the temperature dependence of the density (or specific volume),
the typical time scale of the $\alpha$-relaxation, and the heat capacity.
From this, we argue that the glass transition is dominated by the vicinity of a first order
phase transition.
The fluctuations, observable in principle as scattering, would support the findings and would
increase in terms of amplitude close to the phase boundary (while the size stays constant). 
This amplitude is connected to the cluster size, also introduced in the cooperativity argument.
Minor arguments about corrections from loops are discussed
where we also might have found an argument for the ``Boson Peak''.
The whole concept then bases on equilibrium arguments that are inhibited by -- to our view --
the fluctuations (high susceptibility) plus the high density that results in the strong
growing of the cluster size.
\\[0.5ex]
PACS number(s): 61.43.Gt, 64.70.kj, 64.70.km, 63.50.Lm, 61.25.hk
\end{abstract}

\maketitle

\noindent
The glass transition has been investigated and discussed over many decades with the result of
clear experimental temperature dependencies of the specific volume, the typical
time scale of the $\alpha$-relaxation, and the heat capacity \cite{EJD, jack}.
However, the exact physical meaning of all these observations were left partially unclear such
that different authors even contradict each other \cite{NGAI}.

The scenario becomes even more complicated by the view of Tanaka \cite{tanaka}, who argues that most
glassy systems deal with two order parameters:
One is connected to the particle density as the most important and most obvious oder parameter.
The other one is introduced as a directional order parameter which is connected to
the molecular anisotropy.
For polymers, these fluctuations are connected to the Fischer renormalized behavior \cite{schwahn}
(and references herein).
While spin glasses are dominated by the directional order parameter, classical colloidal suspensions
and simple metals are dominated by the order parameter of the density.

A dynamic model for colloidal suspensions has been developed quite far recently \cite{dhont}.
Astonishingly, hydrodynamic interactions were neglected; thus the model would be applicable
for a wide range of systems.

Here, we offer a model free energy that should help to better understand the experimental observations,
and therefore is able to go a little deeper consistently within the model.
This model considers only density fluctuations, and neglects any directional
order parameter.
Furthermore, it is less heuristic than free volume models \cite{fvol} that successfully
captures most of the features of the glass transition similarly well.
There also exists a lattice theory \cite{freed} that captures aspects of density fluctuations
through the free volume and aspects of directed bonds, but which stays at a finite expansion
level.
We also discuss limits of our model, and accidently came to a motivation of the
``Boson Peak'' observed in \cite{buch}. 

We start with a classical Hamiltonian for a number of $N$ identical particles
that interact with each other via a single pair potential $V(\Delta \vec{r})$:
\begin{equation}
H = T + \sum_{i\neq j}{V(\vec{r}_i - \vec{r}_j)}
\end{equation}
The kinetic energy $T$ is only needed later for the entropy and heat capacity,
and is not of interest for most of the considerations.
The partition function of the interactions will then read:
\begin{eqnarray}
Z_\mathrm{int} = \int_V \!\cdots\! \int_V d^{3N} r_1 \!\cdots\! r_N \exp \left(
-\beta \sum_{i\neq j}{V(\vec{r}_i - \vec{r}_j)} \right. \nonumber \\
\left.
-\frac{\mu}{N-1} \sum_{i\neq j}{\int d^3\Delta r\;\delta(\vec{r}_i-\vec{r}_j-\Delta\vec{r})}
\right)
\end{eqnarray}
We introduced the chemical potential that we will need later for the grand canonical partition.
The potential that is normalized by the thermal energy $\beta^{-1}=k_BT$ can be split
in a short range repulsive and an attractive term of a little longer range, which
is usually considerably strong for next neighbors.
\begin{equation}
\beta V
= v_\mathrm{rep}
+ v_\mathrm{attr}
\approx \epsilon_\mathrm{rep}\delta
+ v_\mathrm{attr}
\end{equation}
We did not write explicitly the argument $\Delta \vec{r}$ for all functions.
This formula and the overall theory will from now on be understood as a
hybrid between a lattice theory and a continuous theory.
The repulsive term reserves a spherical volume of $v_0= \pi d^3/6$ for each particle
of diameter $d$,
and only the nearest next neighbor interactions will be considered.
The term with the chemical potential can be expanded as follows:
\begin{eqnarray}
\!\! & \!\! &
\left\langle
\exp \left(-\frac{\mu}{N-1} \sum_{i\neq j}{\int d^3\Delta r\;\delta(\vec{r}_i-\vec{r}_j-\Delta\vec{r})} \right) \right\rangle \; = \nonumber \\
\!\! & \!\! &
\left\langle 1 -\frac{\mu}{N-1} \sum_{i\neq j}{\int d^3\Delta r\;\delta(\vec{r}_i-\vec{r}_j-\Delta\vec{r})} \right. \nonumber \\
\!\! & \!\! &
\;\; +\frac{1}{2!}\frac{\mu^2}{(N-1)^2} \sum_{i\neq j, k\neq l}{\int d^6\Delta r_1\Delta r_2\;\cdot} \nonumber \\
\!\! & \!\! & \qquad\qquad\qquad\qquad\cdot\;\delta(\vec{r}_i-\vec{r}_j-\Delta\vec{r}_1)\;\delta(\vec{r}_k-\vec{r}_l-\Delta\vec{r}_2) \nonumber \\
\!\! & \!\! &
\;\; \left. -\frac{1}{3!}\frac{\mu^3}{(N-1)^3} \sum_{i\neq j,k\neq l,m\neq n}{\int d^9\Delta r_1\Delta r_2\Delta r_3\cdots}
     \right\rangle \label{corr}
\end{eqnarray}
In this context we used the angled brackets to indicate the canonical integral with the thermodynamic weight of
the original interaction term.
This in particular includes the repulsive term that avoids two particles being at the same origin,
and the next neighbor interactions.
The different lines of Eq.\ \ref{corr} indicate correlation functions with different degrees.
The first line itroduces the 2-point correlation function, the next line a 4-point correlation function, and
so on.
The place in the brackets means, that there are no open ends that are explicitly correlated, but the overall
integral is considered.

Now, we will factorize the correlation functions of the above expression in the following simple manner:
\begin{eqnarray}
Z_\mathrm{int} &=& V^N (1-\phi)^N \sum_{l=0}^{\infty}{\frac{1}{l!} (-\mu N)^l
\left\langle \Phi(\Delta\vec{r}) \right\rangle^l}
\nonumber \\
&=& V^N (1-\phi)^N \exp \Big(-\mu N \left\langle \Phi(\Delta\vec{r}) \right\rangle
\Big)
\end{eqnarray}
The first two factors result from the actual configurationally free volume $(1-\phi)V$ that takes into account
that the particles cannot intersect within the excluded volume. So we define $\phi = Nv_0/V$.
\begin{figure}[ttt]
 \includegraphics[width=5.5cm]{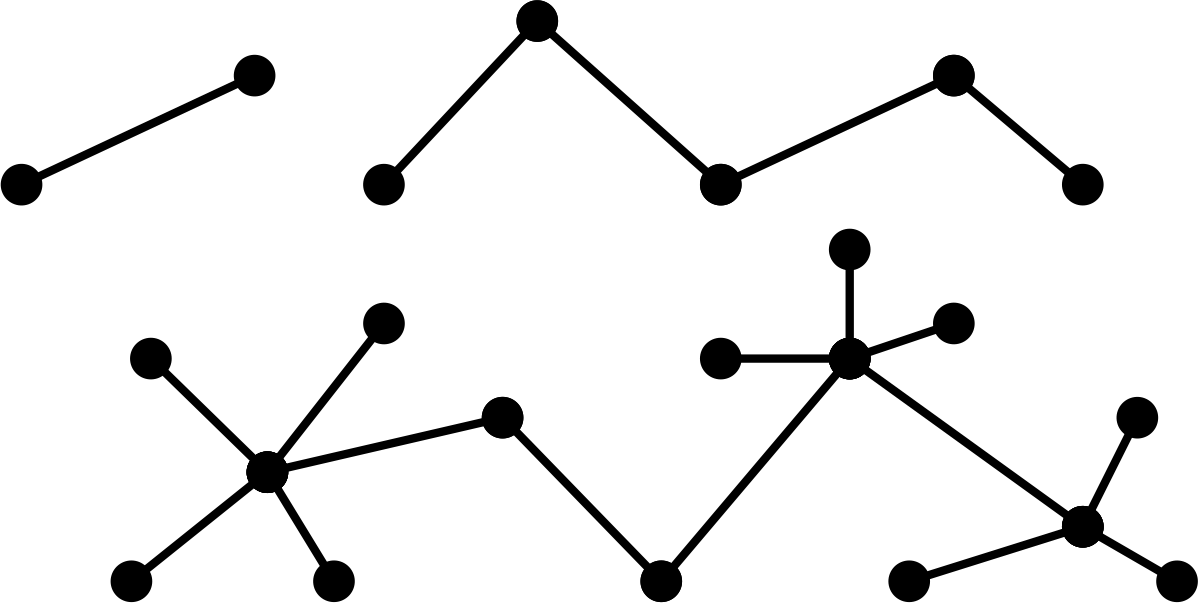}%
 \caption{A selection of open graphs showing the possibility of connectivity.
 None of them violates the factorization of the corresponding correlation function,
 because the order of integration allows for stripping the graphs sequentially.
 \label{graphs}}
 \vspace{-0.45cm}%
\end{figure}%
%
%
%
The factorization of the individual correlation functions is not exact, because various indices might take
the same values which would complicate the correlation function.
First of all, the frequency of indices being different is quite high, especially for the lower degrees
of correlation functions.
Second, the correlations that are represented by open graphs reduce to exactly the factorization, because
the order of integrals can be chosen in this way that the graphs are stripped 
from the outer regions to the inner regions, and each time the exact factor $\Phi$ appears.
So the exactness of this calculation is only corrupted by closed loops (see Appendix A).
The portion of graphs with closed loops is always low compared to the open graphs, especially
for the low degrees of correlations.
Another flaw might be, that at high degrees of connectivity the actually distinguished number of
different paths is limited to the coordination number.
Again, this bulkiness effect will eventually need for corrections at high degrees of correlations, which can
be safely neglected here.
So within the considerations of this manuscript the abovementioned approximation is good,
and assumed to be quite precise.

The derived correlation function is understood as the following:
We consider a chain of particles covering the distance $\Delta\vec{r}$.
From this chain we get a contribution
$\alpha=\phi\cdot(-\epsilon)$ for each particle in the chain accounting for the probability and the energetic
contribution of the next neighbors.
At this point the probability $\alpha$ is treated in the sense of classical lattice theories.
An energy renormalization to the average contact energy would refine the term
$\alpha=\phi\cdot (-\epsilon +\phi\epsilon)=\phi(1-\phi)(-\epsilon)$.
Finally, the correlation function would read then:
\begin{equation}
\Phi(\Delta\vec{r}) = \alpha^n \quad\mathrm{with}\quad n=g|\Delta\vec{r}|/d
\end{equation}
The factor $g$ takes the possibility for wrinkled paths into account, but still is of the order of unity.
The overall estimate is then given by:
\begin{equation}
\left\langle \Phi(\Delta\vec{r}) \right\rangle = 1+\frac{4\pi}{g^3} \sum_{n=0}^{\infty}{n^2 \alpha^n}
= 1+\gamma\cdot\frac{\alpha(1+\alpha)}{(1-\alpha)^3}
\end{equation}
\begin{figure}[ttt]
 \includegraphics[width=6.7cm]{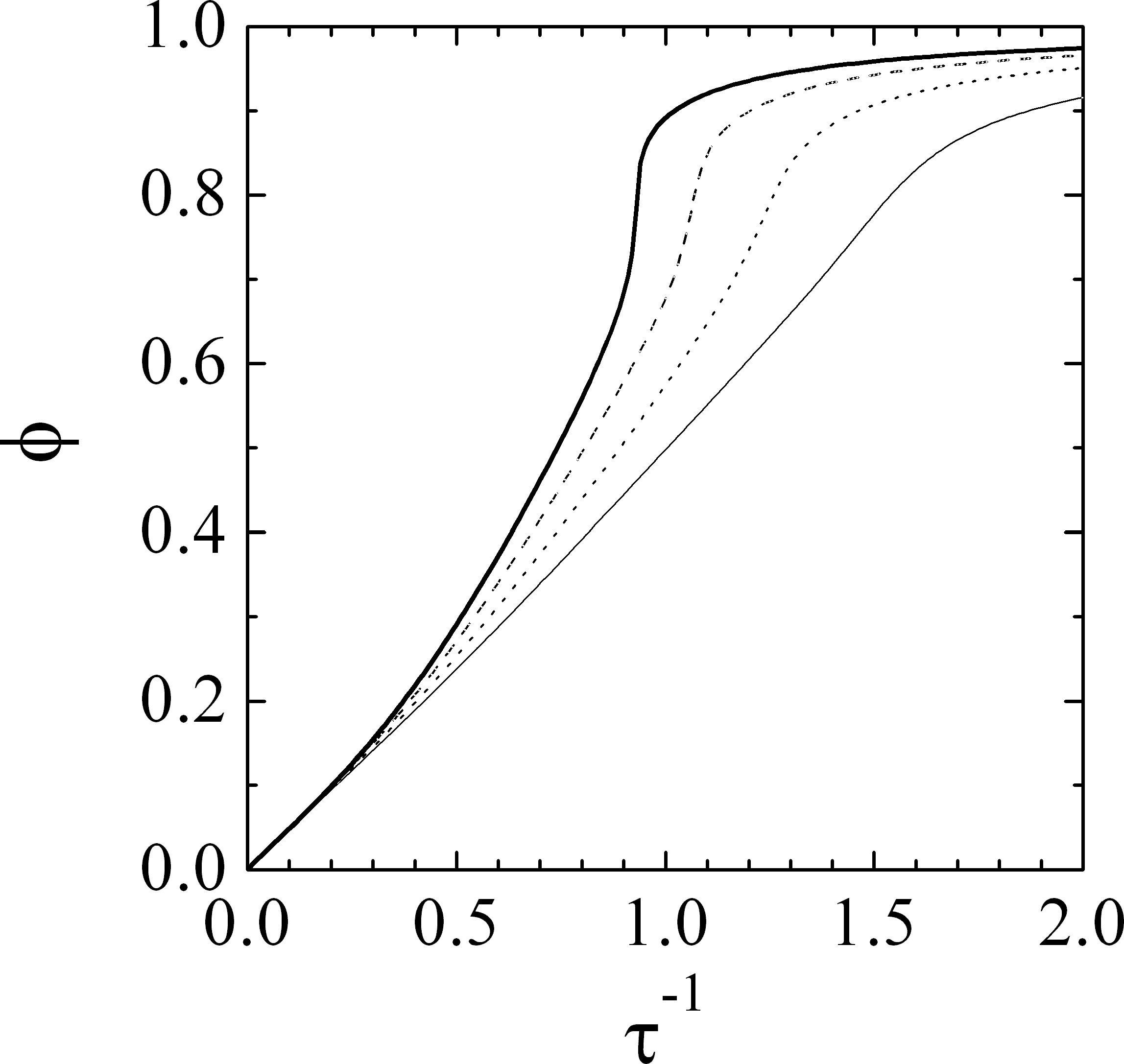}%
 \vspace{-0.3cm}%
 \caption{The particle density as a function of the scaled temperature. The common parameters are
 $\tilde p_0=0.5$ and $\gamma=9.7$. The attractive nearest neighbor interaction $\epsilon_0$ was
 $-0.6$ (solid line), $-0.5$ (dashed), $-0.4$ (dotted), and $-0.3$ (thin solid).
 \label{phi1}}
 \vspace{-0.45cm}%
\end{figure}%
Here we see the hybrid property of continuous and lattice theory, that either 
would yield a prefactor of $4\pi$ or $12$ for hexagonal packed spheres.
Anyhow, we finally include every effect in the prefactor $\gamma$ such as
the coordination number and the wrinkledness of the paths.
It lies in the range of $\gamma=\sqrt{2}\pi=4.4$ and $16\pi/\sqrt{27}=9.7$ for simple cubic
and hexagonal packed lattices.
Finally, we will see below that the exact number does not play a major role.
Apart from the discussion about the coordination number, the explicit presence of the particle
at the origin is added in this correlation function.
This correlation function describes the cluster size with respect to the average
particle density, so to speak the size of enrichments with respect to the background.
%
For the grand canonical partition function, we still need to sum up all possible particle numbers,
and so we obtain:
\begin{equation}
Z_\mathrm{gc} = \sum_{N=0}^{\infty}{Z_\mathrm{int}} =
\frac{1}{1-\left(\frac{V}{V_0}\right)(1-\phi)\exp\Big(-\mu \langle\Phi\rangle\Big)}
\end{equation}
The reference volume $V_0$ is introduced to obtain dimensionless numbers in the partition
function.
It appears as a constant for derivatives and is set to $V$ at the end of calculations.
All of this also guarantees that the abovementioned sum converges to finite numbers.
From this, we do the transformation back to obtain the free energy with the parameters
$T$, $V$ and $N$:
\begin{equation}
F=-k_BT\cdot\frac{N}{\langle\Phi\rangle}\ln\left(\frac{V}{V_0}(1-\phi)\right)
\end{equation} 
This free energy basically looks like a simple expression for free particles that
take a finite excluded volume.
But the number of particles appearing in the expression is now divided by the cluster size $\langle\Phi\rangle$
to yield the effective number of particles.
From this, we obtain the equation of state (with leading terms only):
%
\begin{eqnarray}
-p-\frac{\partial F}{\partial V} &=& 0  \\
\!\!\!\!\!\!
-\tilde{p}\cdot\langle\Phi\rangle + \phi + \frac{\phi^2}{1-\phi}
-\phi^2\frac{\partial\ln\langle\Phi\rangle}{\partial\phi}\ln(1-\phi) &=&0
\label{zustand}
\end{eqnarray}
Here the scaled pressure is $\tilde{p}=pv_0/k_BT$.
This enables us to calculate the particle density as a function of the scaled parameters $\tilde p$
and $\epsilon$.
As all of the parameters scale with the reciprocal temperature $\tau^{-1}$, we can write:
$\tilde{p}=\tilde{p}_0/\tau$ and $\epsilon=\epsilon_0/\tau$.
This basically leaves the interesting parameter space to be around unity for the considered
parameters.

An example for such calculations is shown in Fig.\ \ref{phi1}.
The particle density goes up steadily until $\tau^{-1}\approx 1$, where a steeper increase is found.
We can see that in this example the increase becomes steeper with higher $\epsilon_0$.
As we will see, this property correlates with the fragility of the glass.
For even higher $\tau^{-1}$ the density saturates.
This overall behavior is well known for the specific volume, which is $V/N=v_0/\phi$.
We believe that the theory is closer to the observations on the heating cycle compared to the cooling
cycle, because cooling towards the glassy state deals with stronger changes in the configuration.

\begin{figure}[ttt]
 \includegraphics[width=6.5cm]{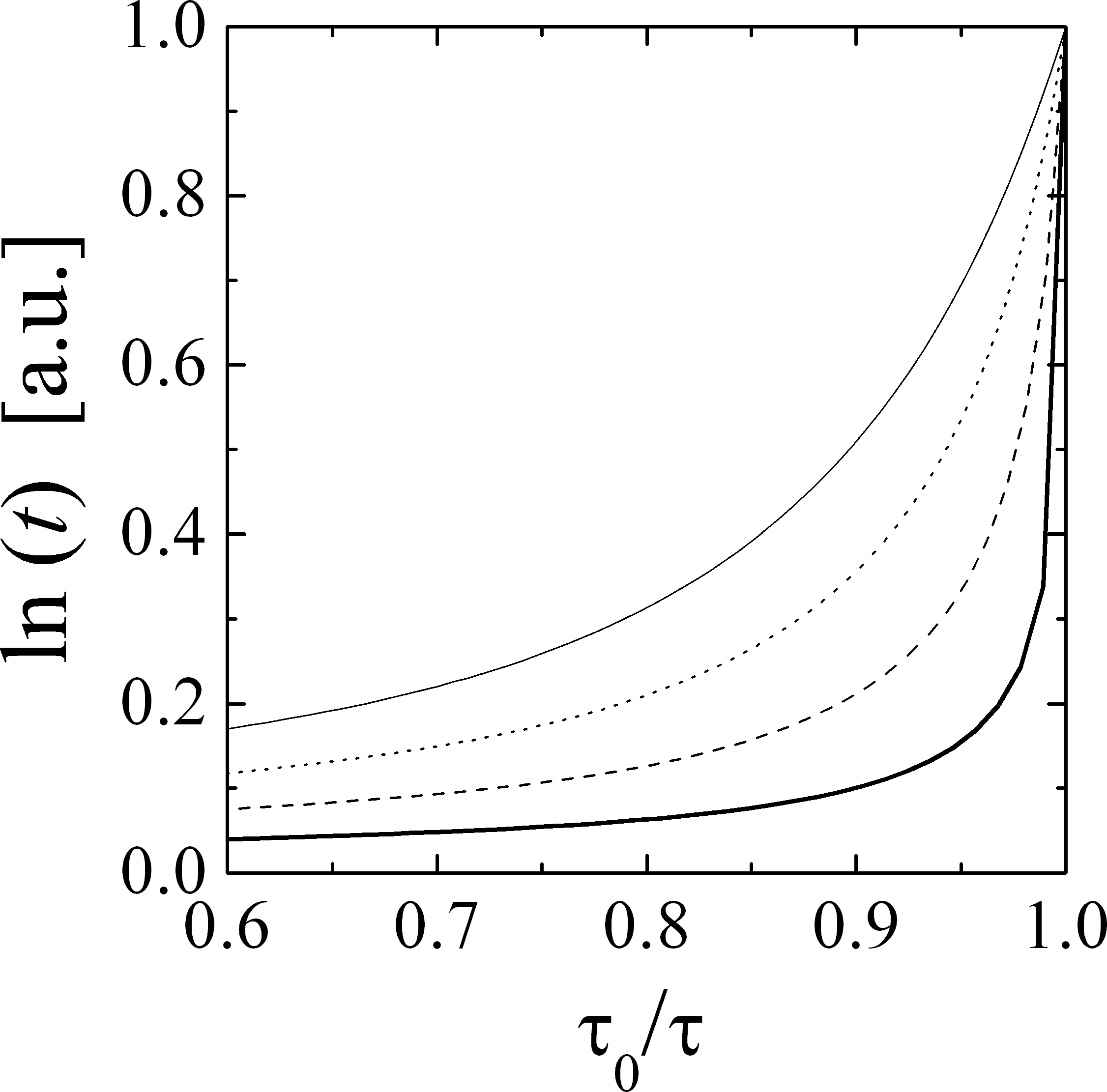}%
 \vspace{-0.3cm}%
 \caption{The characteristic time $t$ as a function of the scaled temperature $\tau_0/\tau$.
 The common parameters are $\tilde p_0=0.5$, $\gamma=9.7$.
 The attractive nearest neighbor interaction $\epsilon_0$ was
 $-0.6$ (solid line), $-0.5$ (dashed), $-0.4$ (dotted), and $-0.3$ (thin solid).
 \label{taus}}
 \vspace{-0.45cm}%
\end{figure}%
From the concept of cooperativity \cite{gibb}, a basic equation was derived for the characteristic time $t$
involved in the $\alpha$-process of the glass forming system \cite{jack, NGAI, gib2, zorn}.
We finally use the specific entropy of the clusters that was explicitly well explained in \cite{gib2}:
\begin{equation}
\ln t = \left( \frac{a}{T\cdot S_\mathrm{conf}} \right)^x
\end{equation}
The parameter $a$ is a free energy and is neglected in the following.
The exponent $x$ is mostly found close to $1$ in theory and experiments.
The configurational entropy is calculated as follows:
\begin{equation}
\frac{S_\mathrm{conf}}{Nk_B}=-\frac{1}{Nk_B}\frac{\partial F}{\partial T}=
\frac{3}{2}+\left(1-
T\frac{\partial\ln\langle\Phi\rangle}{\partial T}\right)
\frac{\ln(1-\phi)}{\langle\Phi\rangle}
\end{equation}
At this point we come back on the kinetic term, i.e.\ $3/2$, that is needed to be considered,
and other terms involving the cluster size.
So, the cluster size is highly important, and gives rise to the glassy behavior,
but is not directly connected to the ``Cooperatively Rearranging Units'' \cite{zorn}, i.e.\ 
$n_\mathrm{cluster}=\langle\Phi\rangle\not\sim S_\mathrm{conf}^{-1}$.

Examples for the relaxation times are depicted in Fig.\ \ref{taus}.
They are normalized to span the range of $0$ to $1$.
The fragile glass ($\epsilon=-0.6$) shows a highly bent curve, while the softer
glass ($\epsilon=-0.3$) leads to a lower curvature.
This is a general observation for glass forming systems, and the underlying model free energy
is capable to connect the fragility of a glass to the simple parameter $\epsilon$:
The nearest neighbor interaction parameter.
These curves can also directly be connected to experimental viscosities \cite{rhe1, rhe2}.

\begin{figure}[ttt]
 \includegraphics[width=6.6cm]{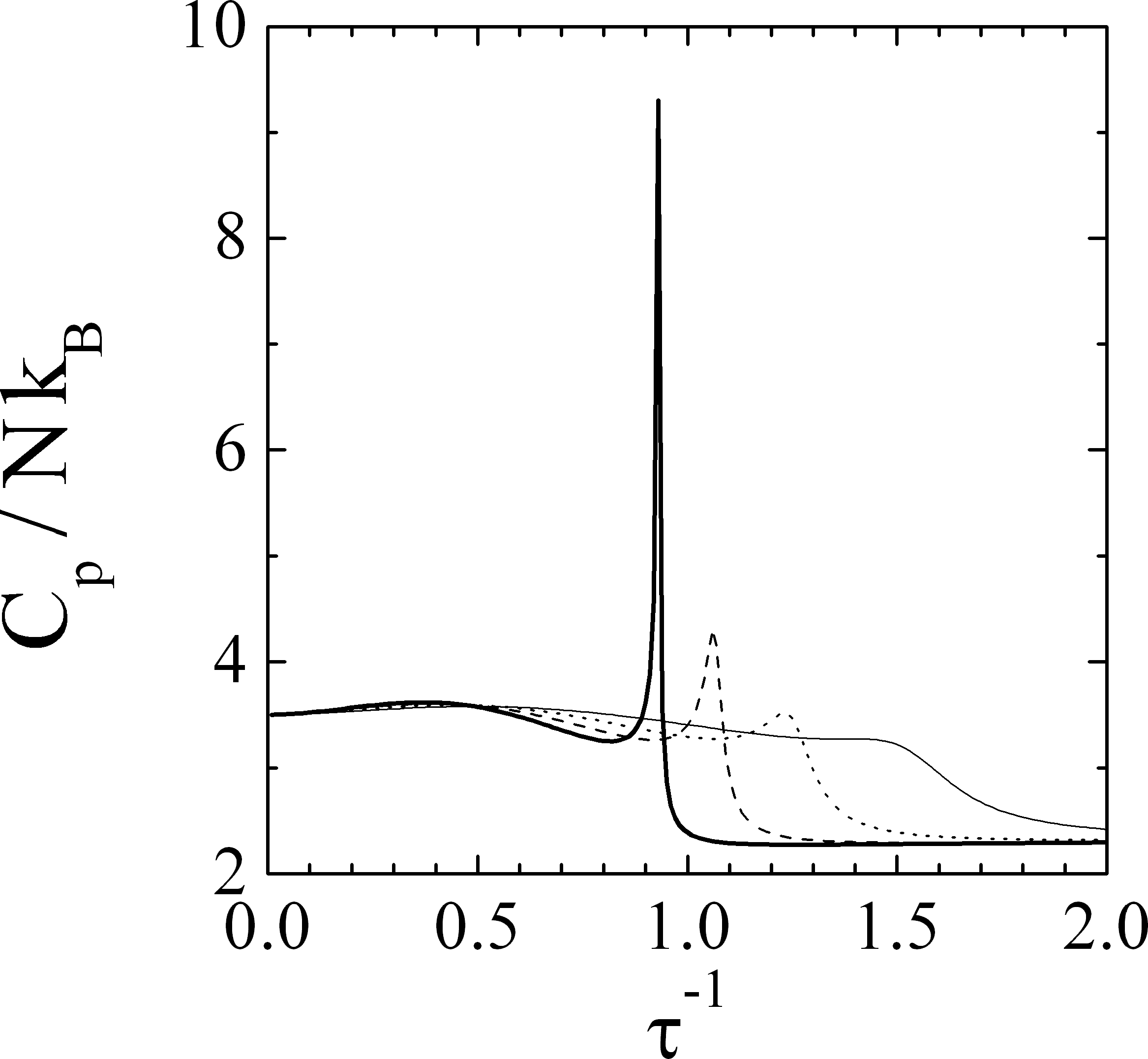}%
 \vspace{-0.3cm}%
 \caption{The heat capacity of a glass forming system as a function of the reduced temperature $\tau$.
 The common parameters are $\tilde p_0=0.5$, $\gamma=9.7$.
 The attractive nearest neighbor interaction $\epsilon_0$ was
 $-0.6$ (solid line), $-0.5$ (dashed), $-0.4$ (dotted), and $-0.3$ (thin solid).
 \label{cpplot}}
 \vspace{-0.45cm}%
\end{figure}%
The heat capacity is another relevant function in concert with the glass transition.
It is related within our expressions in the following:
\begin{eqnarray}
\frac{C_p}{Nk_B}&=&\frac{1}{Nk_B}\left(\frac{\partial H}{\partial T}\right)_p \\
&=&\frac{5}{2}-\frac{T}{Nk_B}\frac{\partial^2F}{\partial T^2}
+\frac{T\tilde{p}}{\phi^2}
\frac{\partial\mathrm{eqn}_{\ref{zustand}}}{\partial T} /
\frac{\partial\mathrm{eqn}_{\ref{zustand}}}{\partial\phi}
\end{eqnarray}
The explicit use of the left hand side of equation \ref{zustand} is used for the full expression.
The kinetic energy also plays a role, and contributes simply in terms of $5/2$.
Examples for the calculation are given in Fig.\ \ref{cpplot}.
We can see the presence of a more or less pronounced peak, according to the fragility of the glass.
We again believe that the theory is closer to the observations on the heating cycle.

Small-angle scattering of glasses would be a measure density fluctuations.
According to the scenario we have developed here, the scattering function is:
\begin{equation}
S(Q) = \frac{v_0\;\phi\;\langle\Phi\rangle}{(1+A^2Q^2/2)^2}
\end{equation}
with the correlation length $A=\sqrt{2}d/(-g\ln\alpha)$.
We expect that especially the correlation length and the specific volume, deliver the
parameters of the presently discussed model.
Details about the coordination number of this model could be corrected
on the basis of scattering experiments on absolute scale.
Furthermore, we believe that the tendency to crystallization might modify the assumed 
coordination number of the hexagonal packing to a different one. 
There are not many experiments on this type of scattering \cite{saxs}, because the correlation
volumes are small, and the scattering intensities are weak.

We just stress the prediction of this model for the scattering:
The correlation length $A$ is rather constant, and the intensity would
increase towards the glass transition.
The regions of stronger fluctuations are stabilized because
a smaller driving force acts towards equilibrium.
This, and the bigger cluster size (or bigger ``Cooperatively Rearranging Units'')
together is responsible for the strong slowing down of relaxations in a glass.
More details about the scattering are discussed in the Appendix B.
%

\begin{figure}[ttt]
 \includegraphics[width=6.5cm]{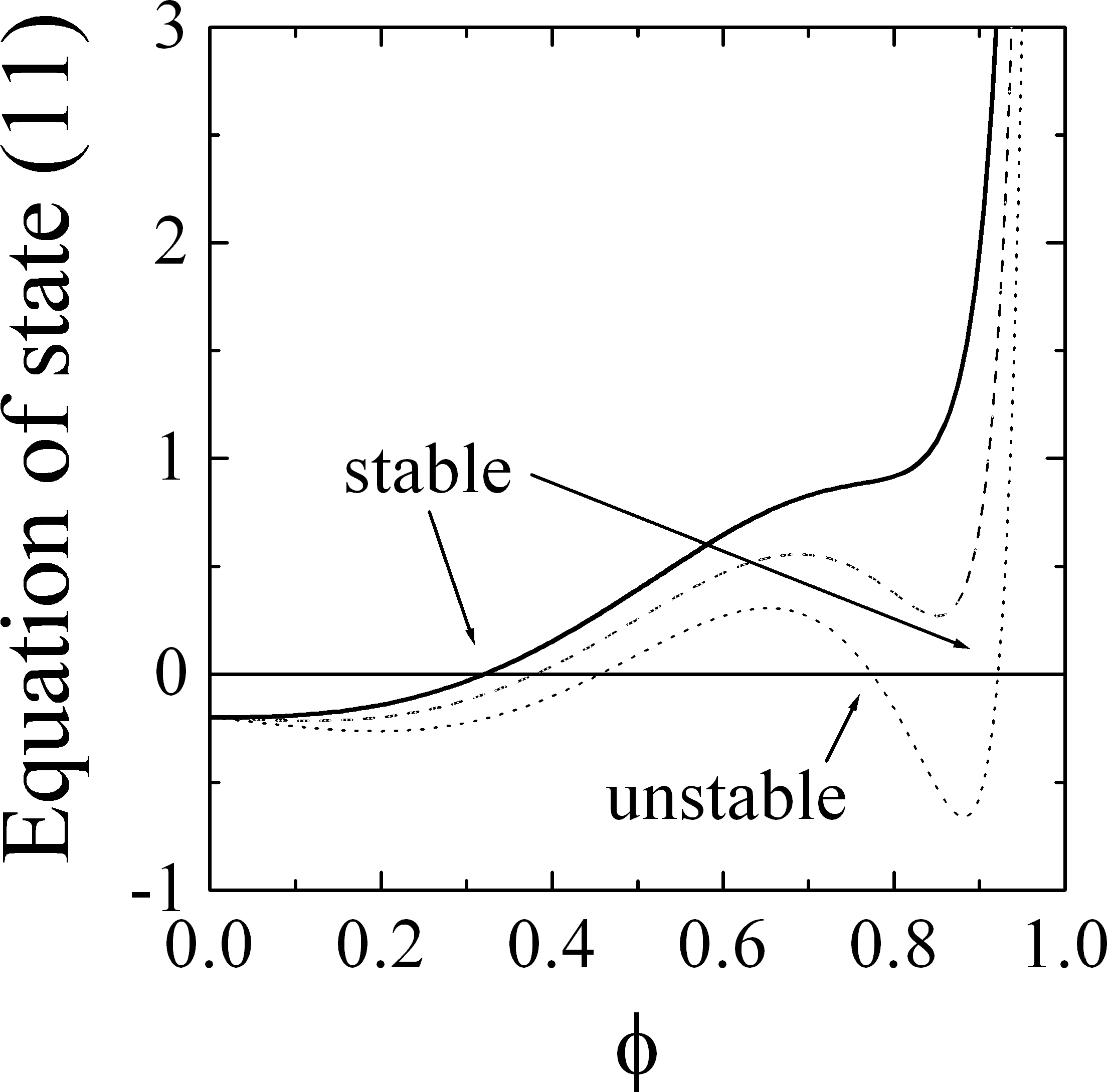}%
 \vspace{-0.3cm}%
 \caption{The equation of state for the parameters $\epsilon=-0.5$ (solid line), $-0.6$
          (dashed), and $-0.7$ (dotted).
         The other parameters were $p=0.2$, and $\gamma=9.7$.
         One can see that a single state is well defined for smaller neighbor interactions,
         while for larger interactions phase separation occurs.
         The interactions for stable phases are indicated by the arrows.
 \label{zustandpic}}
 \vspace{-0.45cm}%
\end{figure}%
We discuss now the meaning of the glass transition within this model.
As we see, that close to $\epsilon_0=-0.6$ the heat capacity diverges, and the specific volume
nearly changes stepwise from one to another value.
We can also depict the equation of state (Eq.\ \ref{zustand}) for higher neighbor interactions
(see Fig.\ \ref{zustandpic}).
We see that then a two-phase coexistence between a high density state and a low density state is
obtained.
While the model is not really capable of predicting the crystalline state correctly,
the meaning is as follows:
The phase transition between the liquid and crystalline state is close to the glass transition.
This causes relatively strong but finite fluctuations of particle density (even though they will be hardly observable
due to a small correlation length).
The discussed equation of state indicates a first order phase transition,
a diverging heat capacity, and a nearly stepwise change of the specific volume.
The interplay of strong fluctuations and large clusters (or ``Cooperatively Rearranging Units'')
explains the dramatic slowing down of relaxations in a glass.
Thus, the glass transition can be classified physically on the basis of this model.
Apart from that, we hope that this model also predicts the glass transition
and its characteristic functions quite quantitatively.

\begin{figure}[ttt]
 \includegraphics[width=6.5cm]{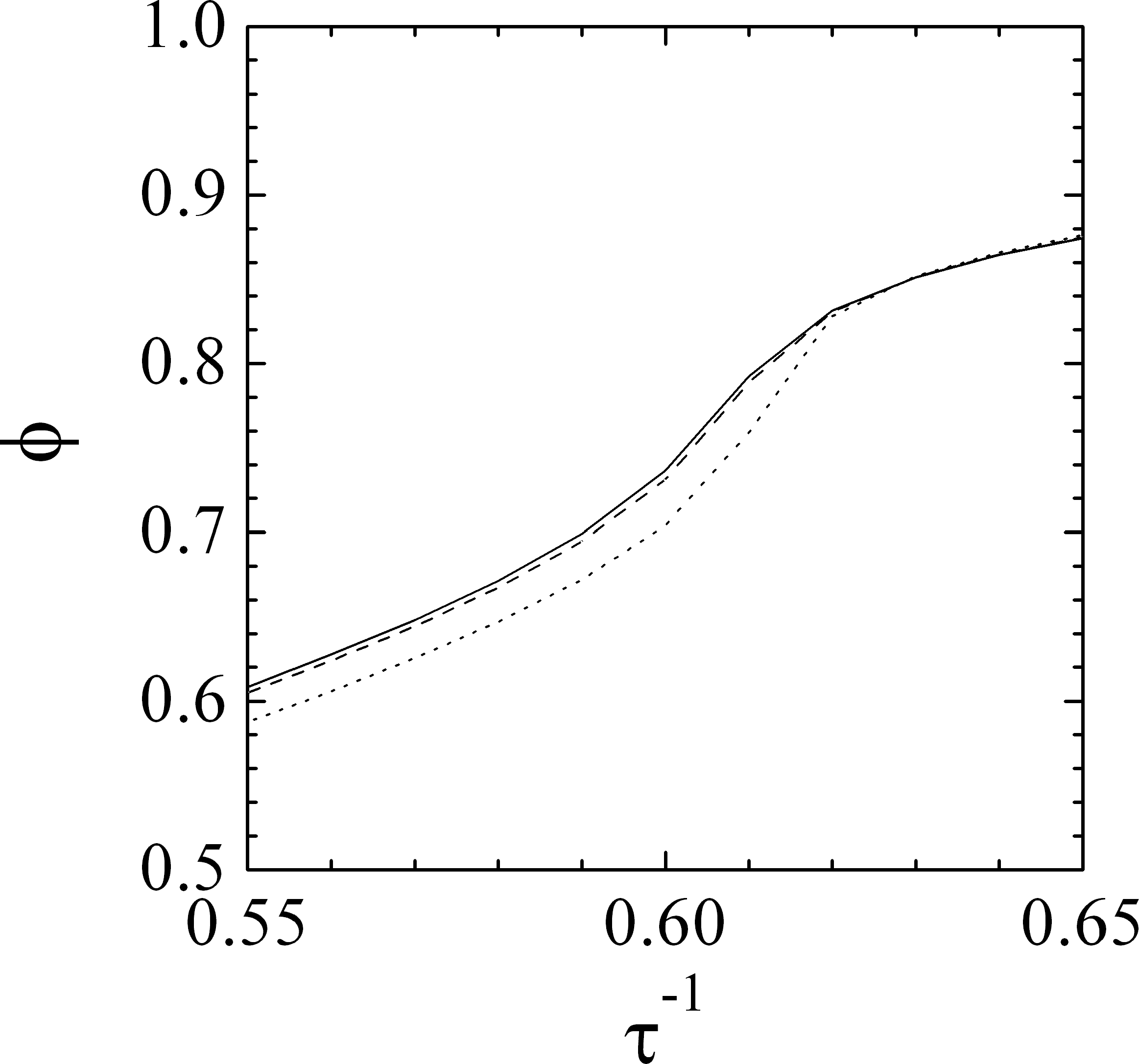}%
 \vspace{-0.3cm}%
 \caption{The particle density as a function of the degree of polymerization,
 which were chosen $N_\mathrm{pol}=500$ (solid line), $100$ (dashed), and $20$ (dotted).
 The common parameters are $\epsilon_0=-1.2$, $\tilde p_0=0.2$, $\gamma=9.7$.
 The overall change of temperature is ca.\ 1\%.
 \label{polym}}
 \vspace{-0.45cm}%
\end{figure}

{\bf The glass transition of polymers} has also been in the focus of recent research  \cite{dgn}.
The whole concept of the present manuscript
might be extended for polymers with little changes (if no directional order parameter
is needed \cite{tanaka,schwahn}).
For the equation of state, we find the following:
\begin{equation}
\tilde{p}\cdot\langle\Phi\rangle =
\phi (1+N_\mathrm{pol}^{-1}) + \frac{\phi^2}{1-\phi}
-\phi^2\frac{\partial\ln\langle\Phi\rangle}{\partial\phi}\ln(1-\phi)
\label{zustandpol}
\end{equation}
The rather strong change lies in the probabilities for the next neighbor particles.
Here the connectivity of the chain plays an important role that is completely independent
of the temperature.
So we find:
\begin{equation}
\alpha = \frac{2}{4\pi}(1-N_\mathrm{pol}^{-1})
+ \frac{4\pi-2(1-N_\mathrm{pol}^{-1})}{4\pi}
\phi(1-\phi)(-\epsilon)
\end{equation}
We calculated some examples for the particle densities as a function of the degree of polymerization
$N_\mathrm{pol}$ (see Fig.\ \ref{polym}).
One can see, that the glass transition temperature is lowered for the shorter chains.
This seems to be the usual case.
Here, the glass fragility seems to be changed less (also seen from the heat capacity, not shown)
compared to smaller molecules.
In parallel, we could find other parameters (smaller $\epsilon_0$, and higher $\tilde p_0$) for
the opposite trend.
While we predicted now changes for polymeric glasses with the degree of polymerization,
a very detailed analysis of the model and experimental data is needed to get the trends correct.
At the moment we leave the results as they are now.

In the Appendix we argue basically for the absence of contributions from loops
in the calculation, because they predict a crowdedness that basically cannot appear.
From calculations of higher order corrections to the scattering and a heuristic
subtraction of the original cluster, we derived an additional scattering function:
\begin{equation}
S_\mathrm{BP}(Q)=
v_0\phi\langle\Phi\rangle^2
\left(\exp(-0.673A^2Q^2)-\exp(-A^2Q^2)\right) \label{bosonp}\\
\end{equation}
This scattering contribution describes a correlation peak,
the intensity of which depends on the magnitude of $n_\mathrm{cluster}$.
Thus, a new length scale $\ell = 5.7A$ appears.
While we believe that this feature would stay weak for a pure structural scattering experiment
(without energy resolution),
the additional correlation might appear in spectroscopic methods,
where this finding is called the ``Boson Peak''.
Polymers might show this feature more often, because the term $\alpha$ includes
a contribution of the connectivity.
The ``Boson Peak'' would describe rearrangements of a few neighbors to the actual cluster.
Details about this feature still stay open for discussion.
We also would like to stress, that we argue here in terms of discretization (or bulkiness) effects
that we introduced heuristically (described by the Feynman graphs in eq.\ \ref{feynmBoson}),
and not in terms of a strict mathematical formalism
(that actually contradicts to this view).

\begin{appendix}
\section{Appendix A: Some missing terms}\label{aa}
The approximations so far left some terms disappear, and we tried to argue that we still cover
most of the physics of the glass transition.
This is quite true, and we will give more details why.
So within second order correlations, a single loop might form. The contribution looks like:
\begin{equation}
\frac{1}{2}\mu^2 \langle\Phi\rangle
\quad\quad\includegraphics[width=1.15cm]{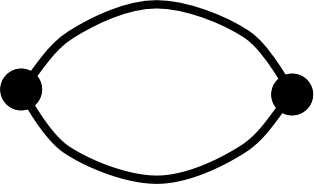}%
\end{equation}
Here and in the following we neglected details about the combinatorial factors, when
$N$ went slowly down to $N-1$, $N-2$ etc.
For the third order correlation we could also have a loop coexisting with a single strand.
This contribution looks similar:
\begin{equation}
-\frac{3}{6}\mu^3N\langle\Phi\rangle^2
\quad\includegraphics[width=2.3cm]{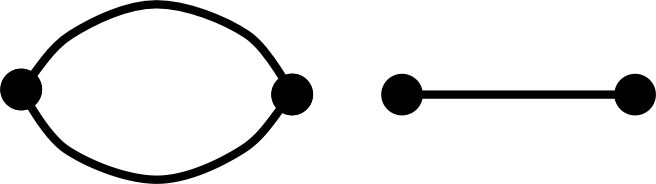}%
\end{equation}
The most complicated correlation we consider is a loop consisting of three strands. The contribution is:
\begin{equation}
-\frac{1}{6}\mu^3 f 
 \left( 1+ \frac{\gamma^{4/3}}{2(2\pi)^{1/3}}\frac{\alpha(\alpha^2+4\alpha+1)}{(1-\alpha)^4} \right)
\quad\includegraphics[width=1cm]{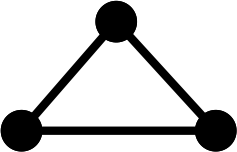}%
\end{equation}
The constant $f=0.01669$ results from a numeric integration for the triangular correlation.
For all sequences of particles we assumed discrete particles, while for the constant $f$
we assumed a continuous geometry first, that was then replaced by a discrete sum to yield the
fraction containing the $\alpha$-terms.

The terms of order $l$ containing a single loop are small compared to the leading term of order $l$
when the condition below is fulfilled:
\begin{equation}
N^2 \gg \frac{l(l+1)}{2\langle\Phi\rangle}
\label{cond1}
\end{equation}
By assuming a large system, this condition is always fulfilled.
The complicated case of small finite volumes \cite{zorn} cannot be treated here because
there are important surface terms apart from the condition above.
The condition for the triangle correlation (here we stick to the 3rd order correlation) reads:
\begin{equation}
N^3 \gg f \left( 1+ \frac{\gamma^{4/3}}{2(2\pi)^{1/3}}\frac{\alpha(\alpha^2+4\alpha+1)}{(1-\alpha)^4} \right)
\langle\Phi\rangle^{-3}
\end{equation}
Again, within our model the system is large, and so the condition is usually fulfilled
(actually more easily than condition \ref{cond1}).
Tending towards smaller volumes would also require the neighbor interaction to be attractive enough
to have a reasonable system within the approximations.

\section{Appendix B: The smallness of the loop terms}\label{ab}
We have seen that at some point the loop terms might break down the approximation.
Here, we would like to argue that a strict mathematical treatment of loops is
to be seen with caution, because those correlations would take place on tyniest spaces,
and that cannot appear for particles of finite size.
While the open loops extend over a reasonable space, the closed loops do not.
So, we discuss a scattering contribution from a triangular correlation, i.e.:
\begin{equation}
v_0\phi\langle\Phi\rangle^2\exp(-1.345q^2)
\quad\quad\includegraphics[width=1cm]{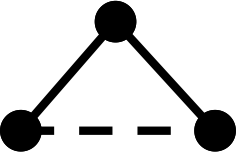}%
\end{equation}
Here we abbreviate $q=dQ/(-g\ln\alpha)$. The dashed line indicates the scattering term.
When comparing this result with the original scattering function, we see:
The structural size does not change dramatically, while the amplitude goes up dramatically.
This would mean, that there are many more correlations in tyniest space.
Due to our applied continuous space description (used especially here), we artifically
find an amplification of correlations for the higher order terms.
This means that closed loops violate the concept of discrete particles.
One way out, to describe higher order terms (heuristically) might be the subtraction
of this crowdedness, i.e. eq.\ \ref{bosonp} and:
\begin{equation}
\includegraphics[width=1cm]{Loop4.png}\quad - \quad
\includegraphics[width=2.3cm]{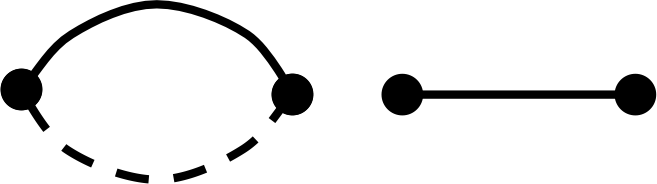}
\label{feynmBoson}
\end{equation}
We stress, that we argue here in terms of discretization (or bulkiness) effects
that we introduced heuristically and not in terms of a strict mathematical formalism.
\end{appendix}

\noindent
{\bf Acknowledgement}:
This manuscript was thoroughly discussed with D.\ Schwahn affiliated with
TUM and Forschungszentrum J\"ulich.


\begin{thebibliography}{99}
\bibitem{EJD} E.J.~Donth, The glass transition: relaxation dynamics in liquids and disordered materials,
              Springer, Berlin / Heidelberg (2013)
\bibitem{jack}J.~J{\"a}ckle, {\it Rep. Prog. Phys.} {\bf 49}, 171-231 (1986)
\bibitem{NGAI}K.L.~Ngai in Soft Matter under Exogenic Impacts, 
              NATO Science Series II: Mathematics, Physics and Chemistry,
              {\bf 242}, 91-111 (2007)
\bibitem{tanaka}H.~Tanaka, {\it J. Chem. Phys.} {\bf 111}, 3163-3174 (1999)
\bibitem{schwahn}D.~Schwahn, V.~Pipich, D.~Richter, {\it Macromolecules} {\bf 45}, 2035-2049 (2012)
\bibitem{dhont}H.~Jin, K.~Kang, K.H.~Ahn, J.K.G.~Dhont, {\it Soft Matter} {\bf 10}, 9470-9485 (2014)
\bibitem{fvol}M.H.~Cohen, G.S.~Grest, {\it Phys. Rev. B} {\bf 20}, 1077-1097 (1979)
\bibitem{freed}K.F.~Freed, {\it J. Chem. Phys.} {\bf 119}, 5730-5739 (2003)
\bibitem{buch}D.~Engberg, A.~Wischnewski, U.~Buchenau, L.~B{\"o}rjesson, A.J.~Dianoux, A.P.~Sokolov,
              L.M.~Torell, {\it Phys. Rev. B} {\bf 59}, 4053-4057 (1999)
\bibitem{gibb}G.~Adam, J.H.~Gibbs, {\it J. Chem. Phys.} {\bf 43}, 139-146 (1965)
\bibitem{gib2}F.W.~Starr, J.F.~Douglas, S.~Sastry, {\it J. Chem. Phys.} {\bf 138}, 12A541/1-18 (2013)
\bibitem{zorn}R.~Zorn, M.~Mayorova, D.~Richter, B.~Frick, {\it Soft Matter} {\bf 4}, 522-533 (2008)
\bibitem{rhe1}S.-E.~Phan, W.B.~Russel, Z.~Cheng, J.~Zhu, P.M.~Chaikin, J.H.~Dunsmuir, R.H.~Ottewill,
              {\it Phys. Rev. E} {\bf 54}, 6633-6645 (1996)
\bibitem{rhe2}J.F.~Brady, {\it J. Chem. Phys.} {\bf 99}, 567-581 (1993)
\bibitem{saxs}T.W.~Wu, F.~Spaepen, {\it Acta metall.} \  {\bf 33}, 2185-2190 (1985)
\bibitem{dgn} B.~Frick, D.~Richter, {\it Science} {\bf 267}, 1939-1945 (1995)
\end{thebibliography}
\end{document}